# Structural and optical properties of high quality zinc-blende/wurtzite GaAs hetero-nanowires

D. Spirkoska[1], J. Arbiol[2], A. Gustafsson[3], S. Conesa-Boj[2], F. Glas[4], I. Zardo[1], M. Heigoldt[1], M. H. Gass[5], A. L. Bleloch[5], S. Estrade[2], M. Kaniber[1], J. Rossler[1], F. Peiro[2], J.R. Morante[2,6], L. Samuelson[3], G. Abstreiter[1] and A. Fontcuberta i Morral[1,7]

[1] Walter Schottky Institut and Physik Department, Technische Universitaet Muenchen
Am Coulombwall 3, 85748 Garching (Germany)
[2] Departament d'Electrònica, Universitat de Barcelona, Marti i Franques, 08028 Barcelona, Spain
[3] Solid State Physics, The Nanometer Consortium, Lund University, Box 118, Lund 22100, Sweden
[4] CNRS-LPN, Route de Nozay, 91460 Marcoussis, France
[5] SuperSTEM Laboratory, STFC Daresbury, Daresbury WA4 4AD, United Kingdom
[6] IREC, Catalonia Institute for Energy Research, Barcelona 08019, CAT, Spain
[7] Laboratoire des Matériaux Semiconducteurs, Ecole Polytechnique Fédérale de Lausanne, 1015 Lausanne, Switzerland

Abstract

The structural and optical properties of 3 different kinds of GaAs nanowires with 100% zinc-blende structure and with an average of 30% and 70% wurtzite are presented. A variety of shorter and longer segments of zinc-blende or wurtzite crystal phases are observed by transmission electron microscopy in the nanowires. Sharp photoluminescence lines are observed with emission energies tuned from 1.515 eV down to 1.43 eV when the percentage of wurtzite is increased. The downward shift of the emission peaks can be understood by carrier confinement at the interfaces, in quantum wells and in random short period superlattices existent in these nanowires, assuming a staggered band-offset between wurtzite and zinc-blende GaAs. The latter is confirmed also by time resolved measurements. The extremely local nature of these optical transitions is evidenced also by cathodoluminescence measurements. Raman spectroscopy on single wires shows different strain conditions, depending on the wurtzite content which affects also the band alignments. Finally, the occurrence of the two crystallographic phases is discussed in thermodynamic terms.

1. Introduction

Heterostructures consist of the combination of two materials, with different band gaps and electron affinities. Heterostructures of type I are formed when a small bandgap semiconductor is surrounded by a larger bandgap material having its conduction-band edge at a higher energy and its valence-band edge at a lower energy than those in the small bandgap material, while heterostructures of type II are characterized by a staggered band-edge lineup. In heterostructures of type I, electrons and holes will tend to localize in the lower bandgap material, while in type II these two types of carriers will be spatially separated. Heterostructures and multilayer-structures have gained huge technological relevance over the past 40 years because the electronic and optical properties of semiconductors can be tailored and modified with respect to their bulk counterpart [1,2]. Heterostructures are formed typically with chemically different semiconductors, such as GaAs and $Al_{1-x}Ga_xAs$. In this work, we present a detailed structural analysis as well as the optical properties of GaAs based 'heterostructures', whose junction is formed by the same material in the two crystalline phases zincblende and wurtzite. Such a combination presents the typical characteristics of heterostructures, because the bandgap and electron affinity of semiconductors depend on the crystalline phase.

In the bulk state, GaAs is stable in the zinc-blende structure. When reduced to a nanoscale volume, such as in the form of nanowire, wurtzite structure becomes stable. In this way, zinc-blende/wurtzite heterostructures in arsenides and phosphides have been realized [3,4,5,6,7,8]. An example of such a wurtzite/zinc-blende/ wurtzite quantum well structure is given in Fig. 1. An aberration corrected high angle annular dark Field (HAADF) scanning transmission electron micrograph (STEM) of a 3 monolayer thick segment of zinc-blende GaAs embedded in a wurtzite GaAs matrix is shown in Fig 1a. A model of the atomic positions has been superimposed. For clarity, this is plotted enlarged in Figure 1b, where the theoretically predicted band alignment [9] of this structure is also superimposed. In such a potential profile, electrons are confined in the zinc-blende segment, while holes occupy the higher valence band states in the surrounding wurtzite structures.

The most successful synthesis method of semiconductor nanowires has made use of the Vapor-Liquid-Solid mechanism, in which the reactants are supplied in the vapor phase and decomposed and/or directly incorporated in a metal seed [10]. The thermodynamics of two-dimensional nucleation of the seeds at the interface between the catalyst droplet and the solid nanowire determines the crystalline quality of the nanowires [11]. In general, the existence of crystal imperfections in semiconductors limits the expected performance of optoelectronic devices. Crystalline imperfections introduced during growth of III-V semiconductor nanowires like InP and GaAs are very common and include rotational twins and polytypism between wurtzite and zinc blende structures [12,13,14,15,16,17,18]. From the crystallographic point of view, zinc-blende and wurtzite structures differ only in the stacking periodicity of the $(111)_{ZB}/(0010)_{WZ}$ oriented planes [19,20,21]. From the fundamental point of view, the occurrence of polytypism in III-V nanowires is particularly interesting. As the wurtzite phase is not stable in the bulk form, the electrical and optical properties are not well known. Twinning or wurtzite/zinc-blende nanowire heterostructures or superlattices should result in a modification of the band structure, generating new forms of band offsets and electronic minibands in a chemically homogeneous material like GaAs [22,23]. From the technological point of view, achievement of control of the wurtzite/zinc-blende phases in nanowires may enable the fabrication of new kinds of devices such as quantum wire cascade lasers [24].

The occurrence of wurtzite structure as one of the main crystalline phases of III-V nanowires has been discussed by several authors. From the thermodynamic point of view, the surface energy of {1100} wurtzite planes might be low in comparison with {110} and {111}A/B of zinc-blende, although detailed atomistic calculations are still lacking. As a consequence, for small radii nanowires, the wurtzite phase would be more stable than the zinc-blende. In the case of GaAs, the critical radius under which wurtzite is expected to be the most stable phase lies between 5 and 25.5 nm, depending on the theory [25,26,27]. Other thermodynamic considerations relate the formation of wurtzite nanowires with a high supersaturation in the catalyst. As nanowires do not grow under thermodynamic equilibrium conditions, kinetic theories have also been used to understand the occurrence of wurtzite nanowires. These theories indicate that the nucleation of new crystalline layers in the nanowire occurs at the triple phase line [20,28]. This type of nucleation kinetically favors crystalline structures exhibiting minimal surface energy facets, again in favor of the wurtzite structure. It is important to note that none of these works refer to nanowire grown without using gold as a catalyst. The optical properties of wurtzite/zinc-blende structures have been investigated in the past by photoluminescence spectroscopy. To date, relatively broad spectra have been obtained, which do not elucidate the quantum nature of these structures [29,30].

In this work, we present structural and optical properties of very clean GaAs zinc-blende/wurtzite nanowire heterostructures with different average contents of wurtzite phases exhibiting localized and sharp emission characteristics. Thanks to the use of local spectroscopy and time-resolved techniques such as confocal photoluminescence and cathodoluminescence, the quantum confined nature of these heterostructures is elucidated. The optical characterization is accompanied with Raman spectroscopy, in order to reveal the presence of strain. Some theoretical considerations on the growth mechanisms of each of the crystalline phases are also included.

2. **Experimental details**

The synthesis of the wires was carried out in a Gen-II MBE system. Two-inch (001) and (111)B GaAs wafers coated with a sputtered 10 nm thick silicon dioxide were used. The nanowire growth was performed at a nominal GaAs growth rate of 0.25 Å/s, a substrate temperature of 630 °C and with 7 rpm rotation. More details on the synthesis process and the growth mechanisms can be found elsewhere [31,32]. In this synthesis method, the arsenic beam flux has been varied from $3.5 \times 10^{-7}$ mbar to $3.5 \times 10^{-6}$ mbar. Here, we present the results on three different arsenic beam flux conditions, which will be referred in the following as samples α, β and γ –see Table 1-. Nanowires of sample α were deposited at $3.5 \times 10^{-6}$ mbar, sample β at $8.8 \times 10^{-7}$ mbar and sample γ at $3.5 \times 10^{-7}$ mbar. By keeping the gallium rate and substrate temperature constant, the arsenic beam flux directly influences the supersaturation of the gallium droplet. In this way, the arsenic beam flux directly determines the growth rate of the nanowires [32]. Nanowires of type α, β and γ correspond to a nanowire growth rate of 3, 1 and 0.3 μm/h, respectively

The samples were prepared for the transmission electron microscopy (TEM) analysis by mechanically removing the GaAs nanowires from the substrate with a razor blade and diluting them in a hexane suspension. A drop was then deposited on a holey carbon copper grid. Before introducing the sample into the microscope, it was introduced into the Plasma Cleaner for 15 seconds. The morphology and structure of the nanowires was characterized by Raman spectroscopy, high resolution transmission electron microscopy (HRTEM) and scanning transmission electron microscopy (STEM) in bright field (BF) and high angle annular dark field (HAADF) modes in a Jeol 2010F field emission gun microscope with a 0.19 nm point to point resolution. Further Cs-corrected HRHAADF analyses were performed on a dedicated VG HB 501 STEM retrofitted with a Nion quadrupole-octupole corrector.

The optical properties were investigated by photoluminescence (PL) and cathodoluminescence (CL) spectroscopy at the single nanowire level. PL spectroscopy experiments on single nanowires were carried out by the use of a confocal microscope embedded in a He$^4$ cryostat [33]. The samples could be scanned below the confocal objective (NA=0.65) with piezopositioners with a spatial accuracy in the nanometer range. The measurements were realized at a temperature of 4,2 K, using the 632.8 nm line of a He-Ne laser as an excitation source. The spot size of the laser in the confocal microscope is about 0.8 μm. The luminescence was detected and analyzed by the combination of a grating spectrometer and Si charge coupled device. The spectrometer has a resolution of 500 μeV. In order to avoid interference of the PL signal with the underlying GaAs substrate, the nanowires were first mechanically removed from the initial substrate and transferred in low densities to an oxidized piece of silicon. Single nanowires could be localized on the surface by scanning reflectivity measurements over areas of up to 30 μm$^2$. Time resolved measurements were realized by exciting with a pulsed Ti:sapphire laser and detecting the photoluminescence intensity as a function of time with a photomultiplier. The laser pulse width and repetition rate were 100ps and 12 ns, respectively.

For CL measurements, the samples were prepared in a similar way on a gold coated silicon substrate – gold ensures a good extraction of the impinging electrons such that no charging occurs-. CL was realized in an adapted scanning electron microscope. The luminescence was detected and analyzed by the combination of a grating spectrometer and a photomultiplier. The temperature of the sample was 9 K. The beam was focused to a small spot of 50 nm in diameter and was scanned over the sample in order to obtain a map of the spectral variations along the sample. More details on the technique can be found in reference [34].

Spatially resolved Raman spectroscopy was realized with a μ-Raman setup in the backscattering configuration on single GaAs nanowires transferred on to a silicon substrate. Prior to the measurements, the nanowires were identified on the substrate by imaging the surface with a camera. The excitation wavelength was the 514.5 nm line or the Ar$^+$ laser. The used laser power of the excitation was about 200 μW (equivalent to 70 KW/cm$^{-2}$), in order to avoid heating of the nanowire [35]. The scattered light was collected by an XY Raman Dilor triple spectrometer with a multichannel charge couple device detector. The sample was positioned on a XY piezostage, which allowed the scanning of the surface (and therefore the nanowire) with a precision of 10 nm.

### 3. Experimental Results

#### 3.1 Crystalline structure

The crystalline structure of the samples type **α**, **β** and **γ** were analyzed by HRTEM and Cs-corrected HRSTEM in HAADF mode. This method has the advantage that the intensity maxima in the micrographs correspond to the atomic positions. Additionally, the observed intensity is nearly proportional to the square of the average atomic number of the elements constituting the atomic column. We observe that nanowires grown under conditions **α** –highest growth rate- are composed of a single zinc-blende structure, while nanowires synthesized under conditions **β** and **γ** have an increasing percentage of the wurtzite phase. Remarkably, the typical zig-zag shape of twinned zinc-blende nanowires was observed in none of the conditions reported here [36]. A typical low resolution TEM measurement of a pure zinc-blende nanowire is shown in Fig.2a. The lateral facets of these nanowires are of the {110} crystallographic family [37]. Some stripes of different contrast with thickness of the order of 100 nm are observed. The periodicity of these zones can vary along the nanowire, and is typically higher for the initial and final stages of growth, as shown in Fig. 2b. These stripes are caused by a rotational twinning, a 180$^o$ rotation around the growth axis. A high resolution TEM analysis of one such twin interface is shown in Fig. 2c.

A quite different structure is observed for nanowires obtained under growth conditions β and γ. For these wires, we observe the existence of different sections with wurtzite and zinc-blende structures. Before we proceed with the statistical analysis of the structure, we present a precise structural analysis of the zinc-blende and wurtzite sections. A high resolution TEM image of the nanowire obtained under conditions β is shown in Fig. 2d. A HAADF image of the same region is superimposed onto the micrograph. Fig 2e shows a magnified image of the boxed region in 2d. A double twinned interface between two wurtzite domains results in a region composed of two zinc-blende unit cells. Twinned planes have been marked with dashed lines, while Ga and As atoms have been marked in orange and green, respectively, for the wurtzite domains, and in red and blue, respectively for the double unit zinc-blende-cell quantum well. As can be seen clearly in the sketch, a 180$^o$ rotational twin in the zinc-blende structure creates the equivalent stacking of a single wurtzite unit in the interface. Two consecutive twins, create an atomic stacking corresponding to 2 wurtzite unit cells, and so on. Following the same model, a 180$^o$ rotational twin in the wurtzite structure would lead to the formation of one GaAs unit cell of zinc-blende -a rotational twin in wurtzite corresponds to a stacking of ABABCBCBCB-. The atomic arrangement in the wurtzite/zinc-blende heterostructure shown in Fig. 2d and e is superimposed in Fig. 2e, so that the different stacking of the planes becomes obvious.

In order to quantify the occurrence of wurtzite structure as a function of the growth conditions, a detailed statistical analysis of many nanowires was realized. In general, the structural properties of the nanowires obtained under the same conditions are very similar. The exact stacking between wurtzite and zinc-blende layers was found to differ, but the average occurrence coincides. Nanowires of type β typically presented four parts, with different ratios of wurtzite-zinc-blende phases. HRTEM micrographs of these four parts are shown in Fig.3. In order to identify the two crystalline phases, we have used a color code. Red denotes wurtzite, while blue and green refer to the two twinned orientations of zinc-blende. In Fig. 3a the structure of top part of the nanowire is presented. Typically, the nanowires grown under these conditions finish with the zinc-blende structure. The density of twins is relatively high, the typical inter-twin distance about 10 nm. Below this zinc-blende structure that has an extension of about one micron, a zone composed of an alternation of zinc-blende and wurtzite domains is observed (Fig. 3b). The domains/regions have a thickness between 5 and 1 nm. This section extends about 0.8 micron on the nanowire. Below this region, the wurtzite sections increase in thickness up to 10 nm and the presence of zinc-blende is gradually reduced as it is shown in Fig. 3 c and d. Overall, the part or the nanowire exhibiting wurtzite structure represents about 30±10% of wurtzite phases.

Nanowires grown under γ conditions exhibit a much higher percentage of wurtzite structure. Typical HRTEM micrographs of the different sections of the nanowire are shown in Fig. 4. There, the growth of the nanowire also ends with a zinc-blende structure. This time the zone with pure zinc-blende structure is only about 50 nm long. Details on the zinc-blende section are shown in Fig. 4b. As shown in the detailed micrographs of Fig. 4 c-e, below this initial section the percentage of wurtzite is very high. There are regions with a high frequency alternation between wurtzite and zinc-blende phases forming a kind of random superlattice and others with extremely thin inclusions of zinc-blende layers in wurtzite (narrow quantum wells). We estimate that the proportion of wurtzite phase under these growth conditions is about 70±10%.

### 3.2 Optical properties

The optical properties of the three types of nanowires were investigated by photoluminescence and cathodoluminescence spectroscopy respectively at 4.2 and 10K. Typical PL measurements realized on single nanowires are presented in Fig. 5. For illustration purposes, the spectra have been shifted vertically by adding an offset. PL spectrum of nanowires type α corresponds to the top spectrum of the figure. As expected, nanowires from sample α exhibit a single PL peak which corresponds to the free exciton luminescence of GaAs at 1.515 eV. The PL characteristics of the other samples are quite different from sample type α –pure zinc-blende. A large downward shift of the photoluminescence and an increased number of sharp peaks are observed for nanowires of samples β and γ. For sample β, several well defined peaks are present at positions between 1.51 and 1.46eV, with a FWHM between 2 and 6 meV with some peaks even much sharper. In the case of sample γ, a similar number of peaks is observed, but downshifted to the energy range between 1.48eV and 1.43eV. No free exciton line of the zinc-blende phase is observed in these wires.

PL measurements have been performed on various other nanowires corresponding to the growth conditions β and γ. The spectra are in all cases very similar. Several luminescence peaks in the ranges 1.51-1.46 for sample β and 1.48-1.43 for sample γ are observed. The exact position and intensity of each peak varies from sample to sample and on the position on the nanowire. These results are in agreement with the fact that the structure varies along the axis of the nanowires. Additionally, the exact sequence of wurtzite and zinc-blende sections varies from wire to wire. In order to illustrate this, scanning confocal photoluminescence measurements of one sample of each of the α, β and γ types are presented in Fig. 6a, b and c, respectively. The PL emission of sample α is rather homogeneous along the wire, with some variations in intensity and linewidth, but altogether consistent with pure zinc-blende GaAs. A single peak at 1.515 eV is observed, which corresponds to the free exciton line of zinc-blende GaAs [38]. Sample β exhibits many more spectral features in the energy range of 1.515 eV and 1.46 eV. The spectral features are not homogeneous along the length. At the top part of the scan, the luminescence is brighter than at the bottom part. The free exciton peak at 1.515 eV is present in only some regions of the nanowire, while lower energy peaks are observed over the whole length. The presence of the various peaks depends on the position on the nanowire. This may be understood by the variation in the structure along the wire as presented in Section 3.1. In the wire of type γ we find that only one part of the nanowire shows luminescence peaks, with spectral features between 1.48 and 1.43 eV. Each of the peaks is located in different regions of the nanowire. It is interesting to note that no PL signal is observed in the top part of the scan up to 1.55 eV – the region between 1.53 and 1.55 eV not shown in the figure-.

Before going into detail of the analysis of the PL measurements, we present the results on cathodoluminescence spectroscopy mapping (Fig. 7). A series of monochromatic cathodoluminescence images of several nanowires of a sample of type β was recorded. These nanowires show an emission pattern at energies between 1.51 and 1.46 eV. Monochromatic images in this energy range exhibit a series of spatially localized bright emission spots, as illustrated in Fig. 7b-4e. These spots appear at different positions along the wire for different energies. By comparing the patterns in the monochromatic images, like in Fig. 7e, we conclude that the different emission energies originate from different locations in the nanowire. By analyzing the intensity profiles, we can conclude that the carrier diffusion length is very short in this region of the wire, less than 100 nm. These characteristics advocate highly for the existence of an array of quantum heterostructures with different quantization energies along the nanowire. The cathodoluminescence of sample α is also provided for reference in Fig. 7g. These nanowires show emission related to the free exciton of zinc-blende GaAs from most of its length. This is in good agreement with the HRTEM measurements and the spatially resolved photoluminescence studies presented above.

## 4. Discussion

### 4.1 Evidence for staggered band-offsets

The photoluminescence peaks and cathodoluminescence maps presented in Fig. 5, 6 and 7 can be interpreted based on the theoretical band gaps and the band alignment in wurtzite/zinc-blende heterostructures [9]. The energy gap of unstrained wurtzite GaAs was predicted to be about 33 meV higher than that of zinc-blende [9]. This has been verified recently with PL measurements of pure wurtzite nanowires [3]. Wurtzite/zinc-blende GaAs heterostructures are believed to exhibit a type II band alignment as shown schematically in Figs. 1b and 8a. Theory predicts conduction and valence band discontinuities of 117 and 84 meV, respectively, where the valence band being higher in wurtzite [9]. Effects of strain and spontaneous polarization at the wurtzite/zincblende interface are neglected here. With this simple model it is expected that an electron hole pair confined to the wurtzite/zinc-blende interface will give rise to a spatially indirect recombination at 1.431 eV -neglecting also the exciton binding energy, which is expected to be small due to the spatially indirect nature of the exciton. This spatially indirect recombination should be the lowest energy observable in pure GaAs wurtzite/zinc-blende multilayer structures. Indeed we observe in wires of type γ the lowest PL line typically at 1.43 eV, in surprisingly good agreement with our simple model. When the thickness of wurtzite and zinc-blende regions are reduced to several nanometers, quantum wells are formed and quantized levels of electrons and holes appear. Some examples of such type of multilayer structures are sketched in Fig. 8a. As a consequence of the type II band alignment, electrons and holes are spatially separated and stored in the zinc-blende and wurtzite regions, respectively. The thinner the quantum well, the higher is the quantization energy and PL lines are expected between 1. 515 and 1.43 eV. In the case where the wurtzite sections are long enough, one would also expect to observe luminescence from this phase at about 1.55 eV [39].That we do not observe this pure wurtzite related luminescence in our samples indicates that the diffusion length of the electrons in the wurtzite segments is long enough that they are trapped in neighboring zincblende regions before they recombine.

We have calculated the first electron and hole confined states of a wurtzite or zinc-blende quantum well embedded in a zinc-blende or wurtzite matrix, respectively. These calculations are indicative and have been realized by assuming the same effective masses for the two GaAs polytypes of electrons and holes, respectively. The energy values are plotted in Fig. 8b. Due to the smaller mass of the electron compared to the hole and the higher conduction band discontinuity, the confinement energy is larger for electrons than for holes. Following these calculations, as the thickness of the quantum wells is reduced to zero, the first level of the electron and hole states tends to the value of the conduction and valence band offsets. Experimentally, we observe that wires of type β show PL lines at higher energies closer to the zinc-blende free exciton line, while PL lines in the lower energy range are observed with increasing wurtzite content (samples γ). This cannot be explained with the simple confinement model presented above. However, one should note that on the nanowire sections where the wurtzite and zinc-blende regions are below ~3 nm, the quantum wells are not isolated but highly frequent. We believe that in these regions, the electron and hole wavefunctions overlap and form minibands that reduce the value of the confinement energy [25,40]. Such kinds of random superlattices are indeed observed in the TEM analysis of the wires of type γ, as shown in Fig. 3b.

Finally, we would like to present further evidence that the observed PL lines are indeed due to spatially indirect transition from confined carriers in type II quantum heterostructures. The lifetime of an exciton is a measure of the recombination mechanism. For a direct exciton recombination in GaAs lifetimes of about a ns or below are expected. On the contrary, spatially indirect recombinations exhibit longer lifetimes due to the reduced overlap of the wavefunctions. In order to further assess the nature of the observed peaks, we have performed time resolved photoluminescence. Results for the 1.515 eV free exciton recombination in a sample of type α and a line at 1.46 eV of sample of type β are shown Fig. 9. The measurements are offset vertically for clarity. The lifetime of the free exciton line at 1.515 eV is below 300 ps –limited by the resolution of the experimental set up-, while the lifetime of the peak at 1.46 eV is on the other hand 8 ns [41]. The lifetime associated with these lower energy peaks is found generally between about 3 and 8 ns. This is in agreement with recent measurements in similar structures [29], and it further supports our interpretation that the lower energy peaks are due to spatially indirect recombinations in type II heterostructures.

### 4.2. Influence of strain on the optical properties

So far we have neglected the influence of strain on the energy gaps and the optical properties of wurtzite/zinc-blende nanowires. It is well known that strain in semiconductors induces changes in the band structure [42]. This manifests itself in a change of the bandgaps, as well as on the band discontinuities in heterostructures. Spatially resolved Raman spectroscopy is a simple technique to obtain information about local strain in semiconductors, as built-in strain leads to a characteristic shift of the phonon frequency [43,44]. We have performed spatially resolved Raman spectroscopy measurements using a μ-Raman set-up on single GaAs nanowires lying on a silicon substrate. We know from atomic force microscopy measurements that the nanowires are lying with one of the facets parallel to the surface. This means that the incident surfaces for the Raman spectroscopy experiments are of the family $\{110\}_{ZB}$ or $\{1010\}_{WZ}$. In backscattering from such surfaces the TO phonon is allowed in first order, the LO phonon forbidden. In zinc-blende

GaAs the TO phonon at the Brillouin zone center has an energy of about 267 cm$^{-1}$, the LO phonon of 291 cm$^{-1}$ at room temperature, respectively [45]. As it has been shown for GaN, The phonon dispersion of the wurtzite phase can be estimated from the zinc-blende by a simple Brillouin zone-folding procedure, as the unit cell of the wurtzite structure is doubled along the (111) direction. Such a model has been used in the case of GaN [46,47]. With this simple procedure we expect a back-folded TO mode close to 250 cm$^{-1}$ in wurtzite GaAs. Raman spectra of nanowires of the type α, β and γ are shown in Fig. 10. In the spectrum of sample type α, only the transversal optical (TO) phonon at 266 cm$^{-1}$ is observed, slightly below the expected bulk value of zinc-blende GaAs. The LO-Mode at 291 cm$^{-1}$ is not observed for these wires, as expected from the selection rules. The TO and weaker LO modes of zinc-blende GaAs are also observed for sample β. In this case the selection rule for the LO mode may be relaxed due to the presence of numerous rotational twins. An additional peak is observed at 255 cm$^{-1}$ that we attribute to the folded TO phonon being at the zone center in the wurtzite structure. This peak position is shifted slightly to higher wavenumbers than what is expected from the back-folding model. For sample γ, we also observe all three peaks. The "zone-folded" TO mode is stronger and it appears at the expected position (~250 cm$^{-1}$), while the other two modes are shifted about 4 cm$^{-1}$ to lower wavenumbers, indicating the existence of tensile strain in the zinc-blende phase. One should note that both the LO and TO are shifted by the same amount, which is expected for small values of strain. The existence of strain in wurtzite/zinc-blende structures can be explained by the difference in lattice constants of the two structures [48]. This means that for zinc-blende rich nanowires the wurtzite phases may exhibit compressive strain (blue shift of the Raman mode) while for wurtzite rich nanowires the zinc blende phases are under tensile strain (red shift of the Raman modes). Details of the Raman studies and the spatial mapping of the wires will be published elsewhere. Here we just want to emphasis that there is evidence for different strain conditions in the wires of type β and γ. It is well known that the bandgaps decrease/increase with tensile/compressive strain, respectively. In addition the band discontinuities vary with strain. Such effects may have a non-negligible influence on the luminescence energies observed in different energy regions for samples of type β and γ.

**4.2 Considerations on the occurrence of wurtzite and zinc-blende phases**

Comparison of nanowires α, β and γ shows that, as the As flux increases, the growth rate increases and the fraction of wurtzite material decreases -similarly, the decrease of the fraction of wurtzite toward the tops of nanowires β and γ is correlated with some increase of the As flux during the growth of each nanowire-. These findings are somewhat surprising since the growth rate is expected to increase with the supersaturation -measured by the difference of chemical potential $\Delta\mu$ between liquid droplet and solid, including the Gibbs-Thomson effect - and since, in Au-catalyzed GaAs nanowires, wurtzite phase has been demonstrated to form at high supersaturation and zinc-blende at lower supersaturation [8].
These seemingly contradictory observations might be reconciled as follows. According to the analysis of Ref. [29], the nucleation rate for each phase $\phi$ is governed not simply by supersaturation but by the nucleation barrier $\Delta G^\phi \propto \gamma_\phi^2 / (\Delta\mu - \Gamma_\phi)$, where $\gamma_\phi$ is the effective edge energy of the 2D nucleus of phase $\phi$ that forms at the nanowire/droplet interface and $\Gamma_\phi$ accounts for the different stacking energies of zb and wz nuclei ($\Gamma_{zb}=0, \Gamma_{wz}>0$). This edge energy is expected to depend on the vapor pressure, in particular when nucleation happens at the triple phase line [28], where part of the edge is directly in contact with the vapor. It might thus happen that, as the As pressure is increased, the supersaturation increases and the nucleation barriers $\Delta G^\phi$ decrease. As a consequence, the growth rate of the nanowire increases, whereas the edge energies of the two phases change. This induces a change of $\Delta G^{wz} - \Delta G^{zb}$ from negative -favoring wurtzite in nanowires γ- to positive -favoring zinc-blende in nanowires α-. This explanation remains speculative, but the high density of phase changes in nanowires β and γ indicate that the present conditions are indeed at the border of the zinc-blende/wurtzite transition. An alternative explanation would be related to the position in which the nucleation events occur. Nucleation at the triple phase line favors wurtzite, while nucleation away from the triple phase line favors zinc-blende. Indeed, as the supersaturation increases and the size of the critical nucleus decreases, the fraction of nucleation events occurring at the triple phase line decreases at the expense of those away from the triple phase line [28,20]. Given the complexity of the system, it is clear that more extended studies on the growth mechanisms need to be done in the future, in order to obtain a more general picture accounting for the existence of wurtzite and zinc-blende III-V semiconductors in the form of nanowires.

## 5. Conclusions

In conclusion, we have synthesized GaAs nanowires by molecular beam epitaxy under three different supersaturation conditions. We have obtained several GaAs polytypes ranging from pure zinc-blende to wurtzite-rich zinc-blende/wurtzite heterostructures, as demonstrated by HRTEM. Photoluminescence spectroscopy studies of these nanowires show sharp lines whose emission energy shifts from 1.515 eV down to 1.43 eV when the percentage of wurtzite material in the nanowire is increased. This is interpreted as evidence for type II band alignment at the wurtzite/zinc-blende interface. The results may open the possibility for structural bandgap engineering of other III-V nanowires and increase the engineering possibilities with nanowires.

*Acknowledgements*
The authors would like to kindly thank T Garma and M Bichler for their experimental support. This research was supported by Marie Curie Excellence Grant 'SENFED', and the DFG excellence initiative Nanosystems Initiative Munich and SFB 631. This work was also supported in part by the Swedish Foundation for Strategic Research (SSF), the Swedish Research Council (VR) and the Knut and Alice Wallenberg Foundation (KAW). The authors would like to thank the TEM facilities in the Serveis Cientificotècnics in Universitat de Barcelona.

**Table 1.** Growth conditions, growth rate and crystalline structure of samples α, β and γ.

| Type | Arsenic beam flux (mbar) | Growth rate (μm/h) | Crystalline structure |
|---|---|---|---|
| α | $3.5 \times 10^{-6}$ | 3 | Zinc-blende |
| β | $8.8 \times 10^{-7}$ | 1 | Zinc-blende with 30 ± 10% wurtzite |
| γ | $3.5 \times 10^{-7}$ | 0.3 | Wurtzite with 30 ± 10% zinc-blende |

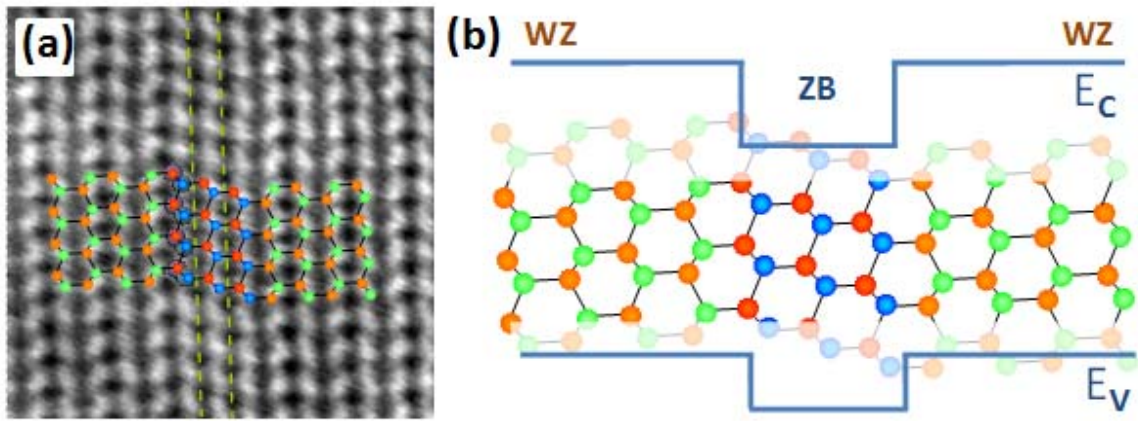

**Figure 1.** Zinc-blende/wurtzite quantum structures. **(a)** Aberration corrected HAADF High Resolution Scanning Transmission Electron Micrograph (HRSTEM) of a zinc-blende quantum well in a wurtzite segment. **(b)** An atomistic model of a wurtzite/zinc-blende/wurtzite heterostructures along with a schematics of the band diagram. The Ga and As atoms have been marked in orange and green, respectively, for the WZ domains, and in red and blue, respectively for the double unit zinc-blende quantum well.

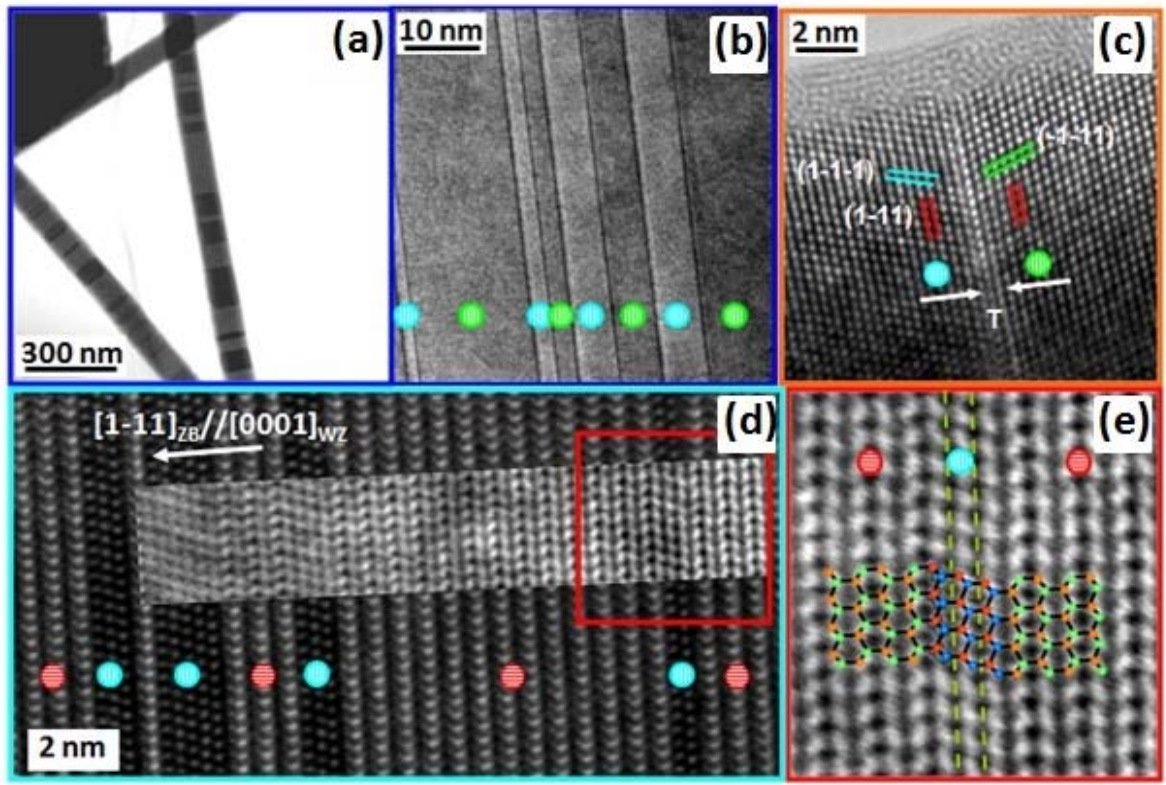

**Figure 2.** Transmission Electron Microscopy of GaAs nanowires presenting different crystalline structures. For illustration purposes, we have indicated each orientation of the zinc-blende phase domains with blue and green circles, and the wurtzite phase regions with red circles. (**a**) General view of the high beam flux nanowires (conditions α) presenting twins in the zinc-blende structure. (**b**) Detail of the twin zinc-blende domains. (**c**) HRTEM analysis of one twin interface. (**d**) HRTEM detail of a nanowire synthesized under conditions β. Inset in (**d**) corresponds to an aberration corrected HAADF HRSTEM image of the same region. (**e**) Magnified Cs-corrected HAADF HRSTEM detail of the squared region in (**d**). A zinc-blende quantum well embedded between two wurtzite regions. Twinned planes have been marked with dashed lines. Simulation of the atomic positions has been superimposed on the measurement.

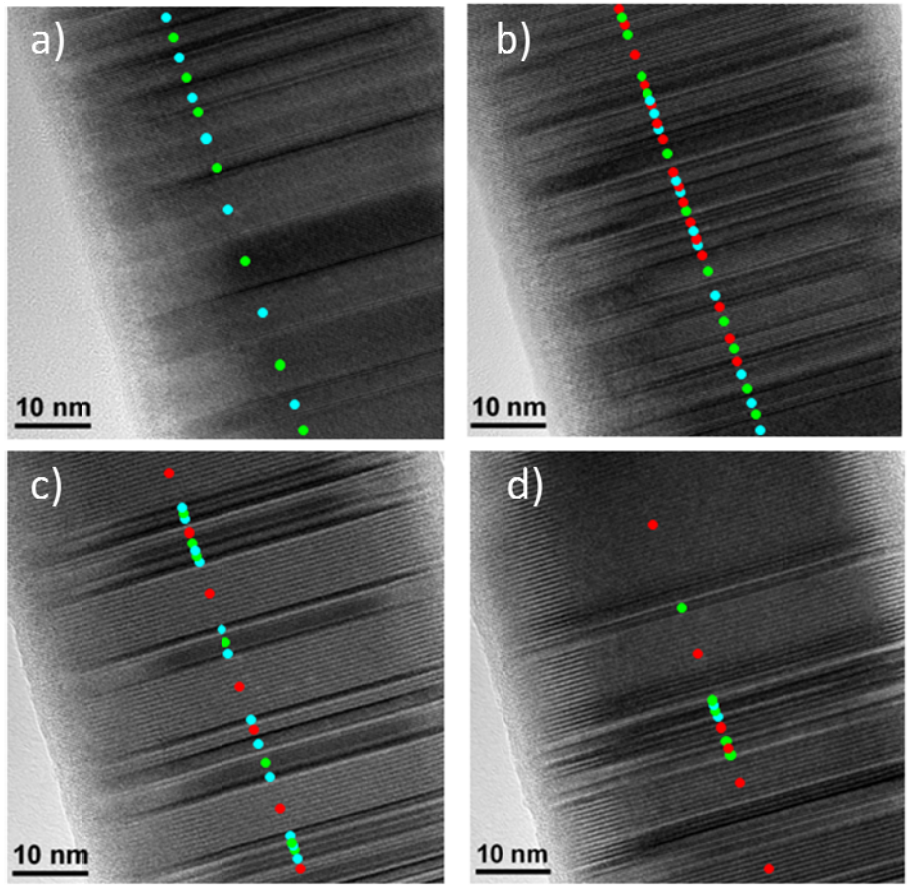

**Figure 3.** Transmission electron micrograph from different segments of a single GaAs nanowires type **β**. From a) to d), the sections belong to different regions of the nanowire from the tip to bottom, corresponding respectively to the latest to initial part of the growth process. The red spots correspond to regions with wurtzite structure, while the blue and green spots correspond to regions with zinc-blend structure –with the two twin orientations-.

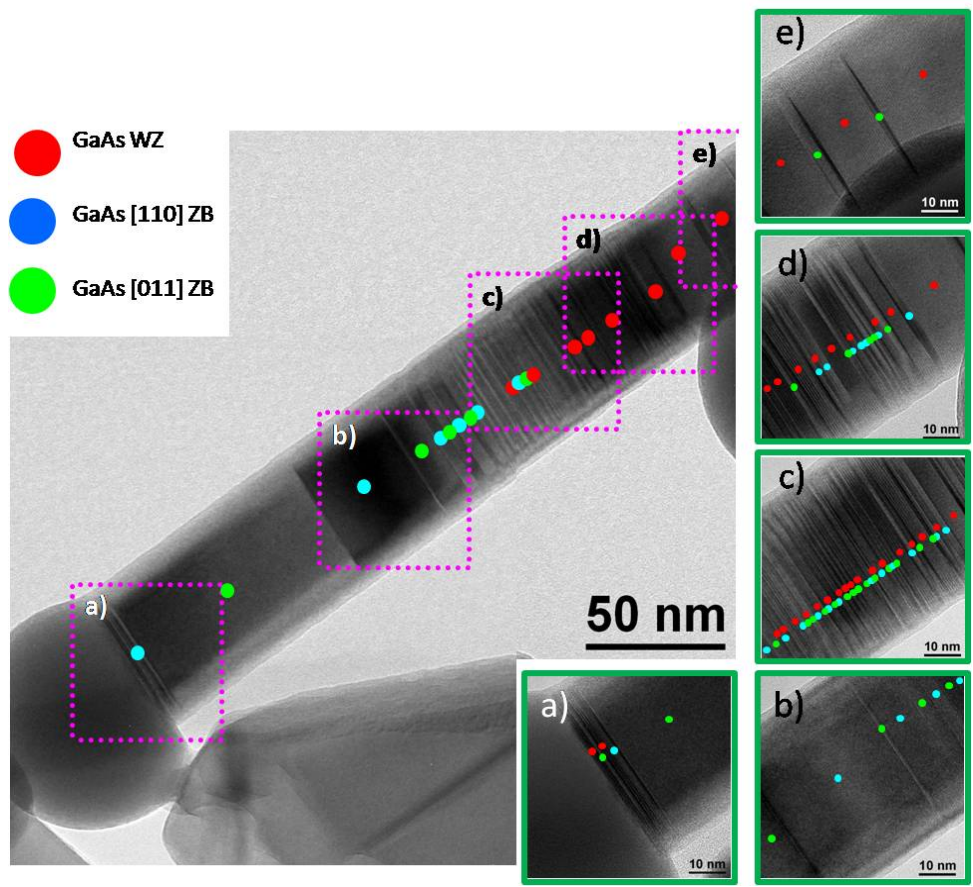

**Figure 4.** Transmission electron micrograph from GaAs nanowire grown under As BF of 3.5x10$^{-7}$ mbar –type γ. The small insets show closer view of the structure from the selected regions.

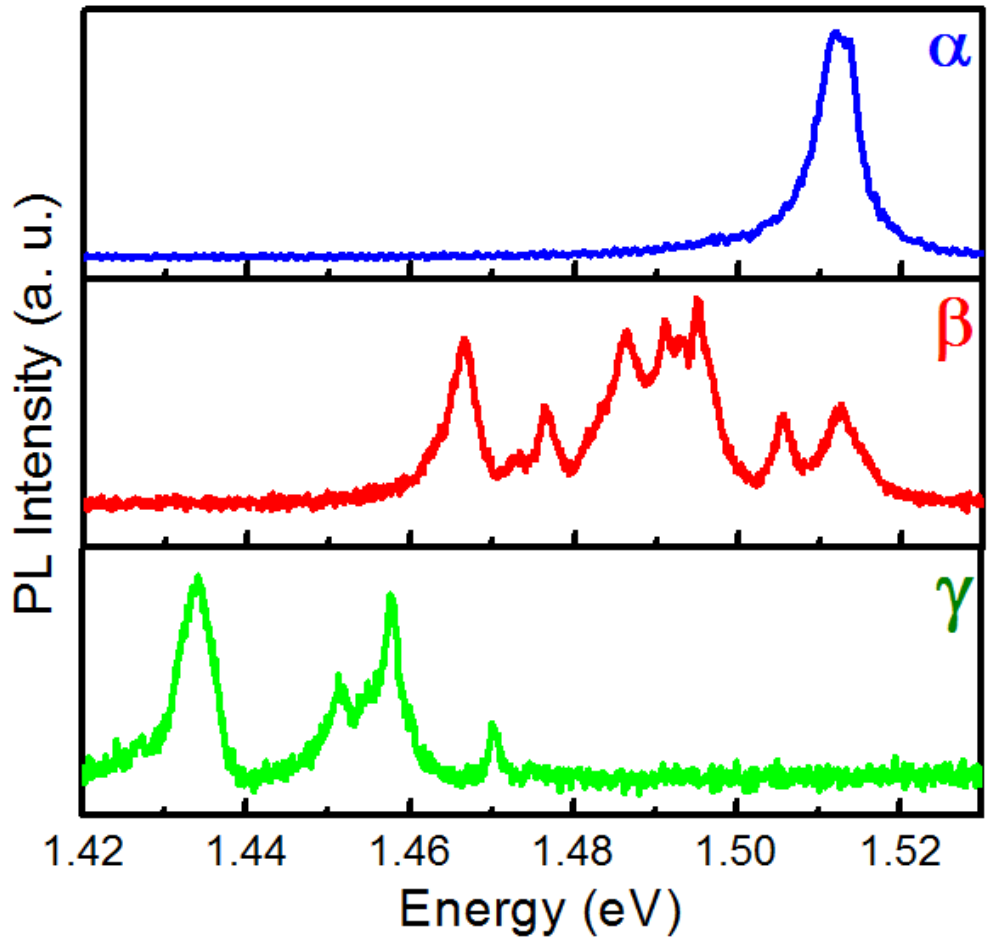

**Figure 5.** Typical photoluminescence spectra observed from the three types of nanowires at T = 4.2 K. The spectra have been obtained from single wires under illumination at 632.8 nm with a power density of of 2.5 W/cm$^2$. The synthesis conditions for samples **α**, **β** and **γ** can be found in Table I. The emission energy shifts from 1.51 eV down to 1.43 eV depending on the growth conditions which result in different proportion of wurtzite and zinc-blende GaAs.

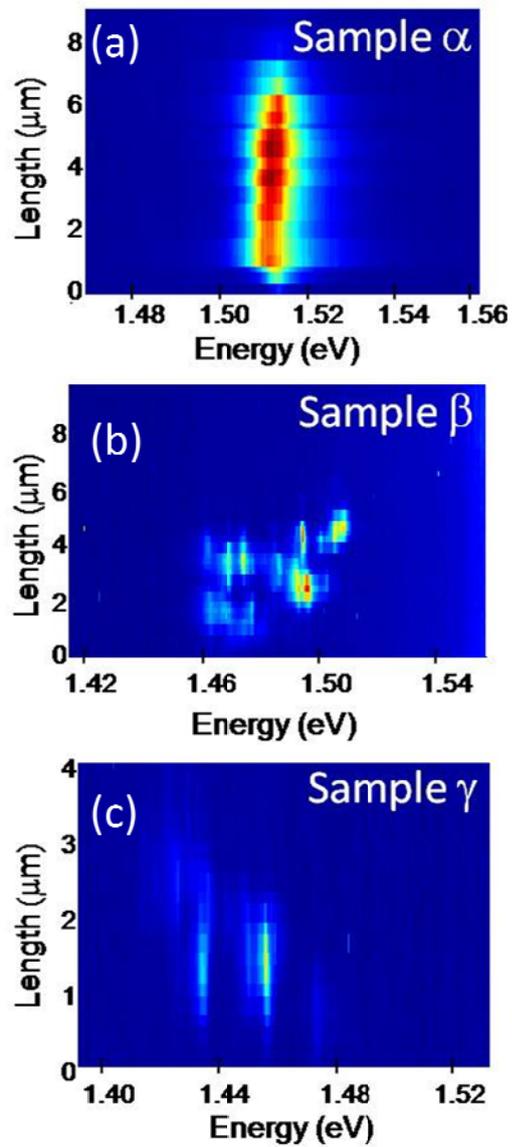

**Figure 6.** Spatially resolved confocal photoluminescence measurements along one single nanowire. The spectra correspond to sample type **α (a)**, **β (b)** and **γ (c).**

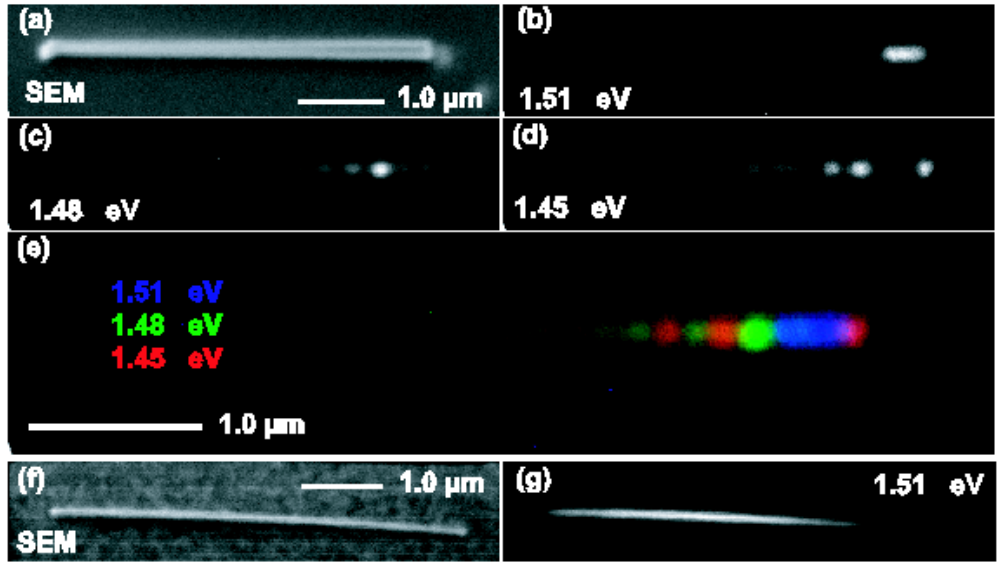

**Figure 7.** Cathodoluminescence of wurtzite/zinc-blende heterostructures. Scanning electron microscopy and monochromatic cathodoluminescence images of nanowires synthesized under conditions α and β. (a) shows an SEM image of a nanowire of type β and (b)-(d) show the corresponding CL images recorded at different energies, illustrating the different spatial origin of the various emission energies. This is even clearer in the composite image of (e). For comparison, we also show images of nanowires type α. (f) is an SEM image and (g) the corresponding 1.51 eV CL image, showing homogeneous emission along the length of the nanowire

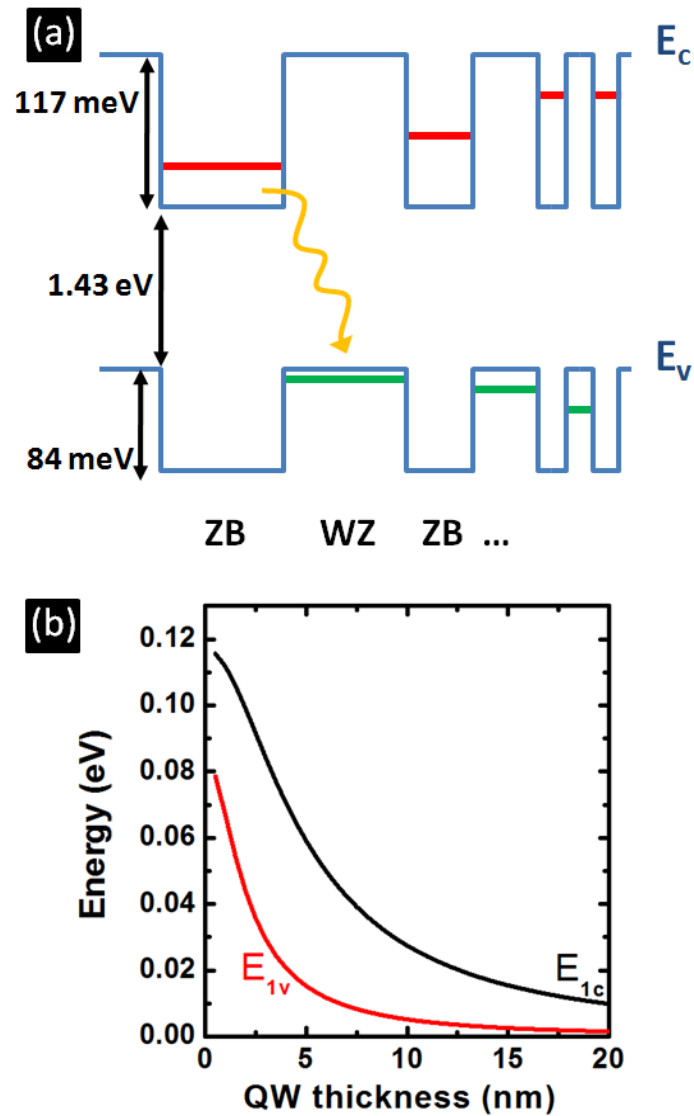

**Figure 8.** Theoretical band alignment of GaAs wurtzite/zinc-blende heterostructures. (**a**). The predicted band alignment of wurtzite and zinc-blende GaAs, along with the values of conduction and valence band discontinuities are shown. Some selected examples of quantum heterostructures which are presented, along with the first quantized level as calculated in (b). The relative scale in energy and length has been kept. The lowest quantized energy levels sketched correspond to quantum wells of 7.5, 5 and 2.5 nm. Due to the type II alignment, the holes are confined in the wurtzite quantum wells, while the electrons are confined in the zinc-blende quantum wells. In (**b**), the energy positions of the lowest electron ($E_{1C}$) and hole ($E_{1V}$) levels of quantum wells as a function of the quantum well thickness are shown.

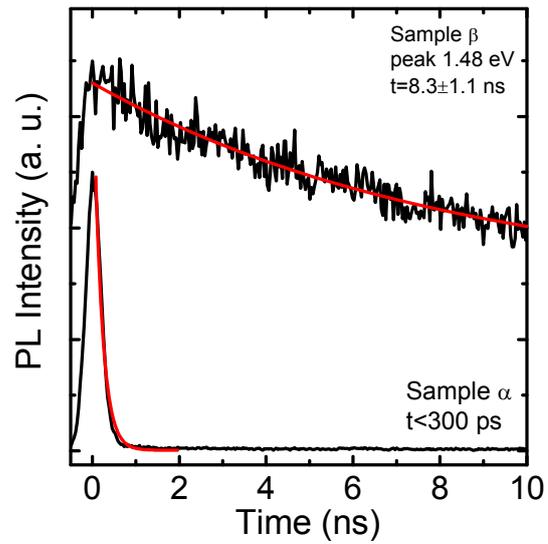

Figure 9. Time resolved luminescence of the free exciton peak at 1.515 eV -sample **α** -and the peak at 1.46 eV from sample type **β**. The measurement curves are separated vertically for clarity.

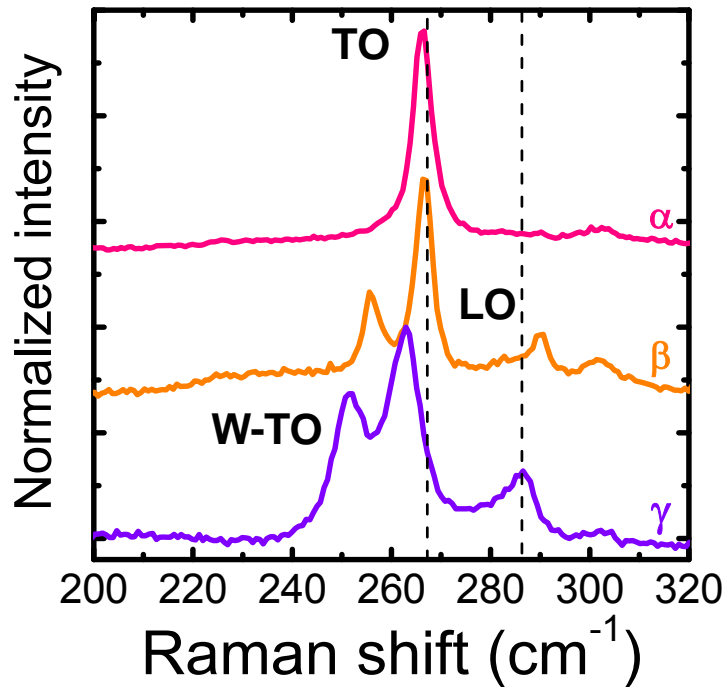

**Figure 10.** Typical Raman spectra of single GaAs nanowires transferred on a silicon substrate. The spectra belong to the three types of nanowires synthesized at an Arsenic Beam Flux of $3.4\times10^{-6}$ mbar (α), $8.8\times10^{-7}$ mbar (β) and $3.5\times10^{-7}$ mbar (γ). In sample α only the TO mode is observed, which corresponds to what is typically observed in backscattering from (110) zinc-blende GaAs surfaces. The LO phonon appears weakly for sample (β), along with a mode at lower frequencies which corresponds to the "folded" wurtzite mode. In sample γ, this mode is stronger and shifted downwards. Note that also the zinc-blende related modes are shifted to lower wavenumbers which is believed to be caused by different strain situations due to different contents of wurtzite phases.